\documentclass{emulateapj}
\usepackage{color}
\usepackage[below]{placeins}
\usepackage{natbib}
\slugcomment{Accepted for publication by the Astrophysical Journal}

\def\psd{phase-space density}
\def\gtorder{\mathrel{\raise.3ex\hbox{$>$}\mkern-14mu
    \lower0.6ex\hbox{$\sim$}}}
\def\ltorder{\mathrel{\raise.3ex\hbox{$<$}\mkern-14mu
    \lower0.6ex\hbox{$\sim$}}}
\def\hmpc{h^{-1} \mathrm{Mpc}}

\def \etal {{\it et al.}}

\def \eg{{\it e.g.,}}

\def\calQ{{\mathcal Q}}

\newcommand {\rs} {$R_{\rm s}$}
\newcommand {\ms} {$M_{\rm s}$}

\newcommand {\rvir} {$R_{\rm vir}$}
\newcommand {\mvir} {$M_{\rm vir}$}

\newcommand {\LCDM} {$\Lambda$CDM}

\shorttitle{Evolution of  phase-space density in dark matter halos}
\shortauthors{Hoffman et al.}

\begin{document}

\title{Evolution of the Phase-Space Density in Dark Matter Halos}

\author{
Yehuda Hoffman\altaffilmark{1},
Emilio Romano-D\'{\i}az\altaffilmark{2},
Isaac Shlosman\altaffilmark{2},
Clayton Heller\altaffilmark{3}
}
\altaffiltext{1}{
Racah Institute of Physics, Hebrew University; Jerusalem 91904, Israel
}
\altaffiltext{2}{
Department of Physics and Astronomy, 
University of Kentucky, 
Lexington, KY 40506-0055, 
USA
}
\altaffiltext{3}{
Department of Physics, 
Georgia Southern University, 
Statesboro, GA 30460, 
USA
}

\begin{abstract}
The evolution of the phase-space density profile in dark matter (DM)
halos is investigated by means of constrained simulations, designed to
control the merging history of a given DM halo. Halos evolve through a
series of quiescent phases of a slow accretion intermitted by violent
events of major mergers. In the quiescent phases the density of the
halo closely follows the NFW profile and the phase-space density
profile, $Q(r)$, is given by the Taylor \& Navarro power law,
$r^{-\beta}$, where $\beta\approx 1.9$ and stays remarkably stable over
the Hubble time.  Expressing the phase-space density by the NFW
parameters, $Q(r)=Q_s (r/R_s)^{-\beta}$, the evolution of $Q$ is
determined by $Q_s$.  We have found that the effective mass surface
density within $R_s$, $\Sigma_s\equiv \rho_s R_s$, remains constant
throughout the evolution of a given DM halo along the main branch of
its merging tree. This invariance entails that $Q_s \propto
R{_s^{-5/2}}$ and $Q(r) \propto \Sigma{_s^{-1/2}} R{_s^{-5/2}} \Big(r
/ R_s\Big)^{-\beta}$.  It follows that the phase-space density remains
constant, in the sense of $Q_s=const.$, in the quiescent phases and it
decreases as $R{_s^{-5/2}}$ in the violent ones.  The physical origin
of the NFW density profile and the phase-space density power law is
still unknown. Yet, the numerical experiments show that halos recover
these relations after the violent phases. The major mergers drive
$R_s$ to increase and $Q_s$ to decrease discontinuously while keeping
$Q_s \times R{_s^{5/2}} = const$. The virial equilibrium in the
quiescent phases implies that a DM halos evolves along a sequence of
NFW profiles with constant energy per unit volume (i.e., pressure)
within $R_s$.

\end{abstract}

    
\section{Introduction}
\label{sec:intro}

The dynamics of dark matter (DM) halos in the Cold Dark Matter (CDM)
cosmology can be easily formulated as the classical Newtonian $N$-body
problem. Yet, the understanding of the equilibrium configuration of
the DM halos defies a simple analytical approach. The lack of
analytical understanding is often compensated for by numerical
simulations which provide empirical knowledge. The cumulative work in
cosmology over the last decade or so has led to a broad consensus
about two basic facts that describe the equilibrium structure of DM
halos.  One is that the spherically-averaged density profile $\rho(r)$
is well approximated by the so-called NFW profile (\citeauthor{nfw96}
\citeyear{nfw96}, \citeyear{nfw97}) or some close variants of it
(\citeauthor{moore98} \citeyear{moore98}; \citeauthor{jing00}
\citeyear{jing00}; \citeauthor{kly01} \citeyear{kly01}). The other is
the power law behavior of the phase-space density profile, namely
$Q(r)=\rho(r)/\sigma^3(r) \propto r^{-\beta}$, with $\beta \approx
1.9$, where $\sigma(r)$ is the velocity dispersion (\citeauthor{tay01}
\citeyear{tay01}).

Two seemingly orthogonal approaches to the problem of the origin of
the equilibrium structure in DM halos exist. One assumes a monolithic
collapse of a halo that can be approximated by the spherical infall
model \citep{gg72}. The application of the model to the cosmological
context, where the shell crossing has to be explicitly accounted for,
has resulted in the so-called secondary infall model (SIM;
\citeauthor{gunn77} \citeyear{gunn77}; \citeauthor{fg84}
\citeyear{fg84}; \citeauthor{bert85} \citeyear{bert85};
\citeauthor{hs85} \citeyear{hs85}; \citeauthor{rg87} \citeyear{rg87};
\citeauthor{zb93} \citeyear{zb93}; \citeauthor{nusser01}
\citeyear{nusser01}; \citeauthor{lh00} \citeyear{lh00}).  The SIM has
been tested against the $N$-body simulations and has proven to
faithfully reproduce the density profile of simulated DM halos
(\citeauthor{qsz86} \citeyear{qsz86}; \citeauthor{efwd88}
\citeyear{efwd88}; \citeauthor{cro94} \citeyear{cro94};
\citeauthor{asc04} \citeyear{asc04}; \citeyear{asc07}).  A closely
related variant of the SIM replaces its dependence on the primordial
over-density of the proto-halo by the mass accretion history (MAH) of
the halo (\citeauthor{ns99} \citeyear{ns99}; \citeauthor{lu06}
\citeyear{lu06}; \citeauthor{ss07} \citeyear{ss07}). The SIM and its
MAH variant can reproduce also the power law behavior of the \psd\
(\citeauthor{au05} \citeyear{au05}; \citeauthor{gon07}
\citeyear{gon07}).

However, a close inspection of the $N$-body simulations reveals that a
DM halo evolves very differently from a monolithic quasi-spherical
collapse. In fact halos are numerically observed to go through a
sequence of mergers, some labeled as major mergers in which the two
main progenitors are of a similar mass, leading to emergence of the
NFW density profile (\citeauthor{sw98} \citeyear{sw98};
\citeauthor{dek03} \citeyear{dek03}; \citeauthor{sco00}
\citeyear{sco00}). \citeauthor{rd06} (\citeyear{rd06};
\citeyear{rd07}, hereafter Paper~I and II) studied the formation and
equilibrium configuration of halos by means of controlled $N$-body
simulations, with the initial conditions set by constrained
realizations of Gaussian fields. These simulations were designed to
address issues of how the merging history affects the DM halos. The
emerging picture is that of a halo evolving {\it via} a sequence of
quiescent phases of a slow mass accretion intermitted by violent
episodes of major mergers.  In the quiescent phases, the density is
well fitted by an NFW profile, the inner (NFW) scale radius \rs\ and
the mass enclosed within it (\ms) remain constant, and the virial
radius (\rvir ) grows linearly with the expansion parameter ($a$).  In
the violent phases, the halos are not in a dynamical equilibrium, but
are rather in a transition state, resulting in a discontinuous growth
of \rs\ and \rvir. In such a picture a halo is defined in the context
of a merger tree --- at any given time it is taken as the most massive
progenitor along the branch leading to the final halo.

A direct comparison between the SIM and numerical simulations has been
conducted recently by \citet{asc07}. This comparison is based on
selecting the DM halos from a cosmological simulation, tracing them
back in time, and recovering their individual initial conditions.  The
SIM has been applied to the `primordial' density profiles, and their
virial density profiles have been calculated for different
redshifts. The SIM calculated profiles provided a good match to the
evolution and structure of the simulated clusters. This is
encouraging.  However, a rigorous fundamental theory that can
accommodate both the spherical monolithic collapse and the major
merger-driven evolution exhibited by the simulations is still missing.
This motivates us to look further into the phenomenology of the \psd,
to gain a further insight into this seemingly simple, yet complicated,
problem.

The study of the DM halo evolution has been heavily focused on the
density profile, while the evolution of the \psd\ has been largely
ignored. \citet{pei07} presented one of the few studies of $Q(r)$
evolution. They found that $Q$, defined as a global quantity
characterizing the halo as a whole, is generally decreasing with
time. Specifically, it exhibits an early rapid decrease (at redshifts
$z \gtrsim 6.5$) and a late slow decrease. Here we aim at studying the
evolution of the \psd\ within the framework of the NFW scaling and the
dynamical picture formulated in Papers I and II. In particular, we
shall rely on the empirical fact that the dynamics of the DM halos is
constrained by a new invariant of motion. This invariant tags a halo
along the main branch of the merging tree. The NFW scaling and this
invariant of motion provide a full description of the cosmological
evolution of the \psd\ for individual halos.

The structure of the paper is as follows. The analysis of the DM halos
in Papers I and II is briefly reviewed in \S \ref
{sec:numerics}. General considerations of the evolution of DM halos
are given in\S \ref{sec:general}. The general evolution of the DM
\psd\ profile is described in \S \ref{sec:results} and self-similarity
and scaling relations are given in \S \ref{sec:scaling}.  A general
discussion follows in \S \ref{sec:disc}.


\section{Numerical experiments}
\label{sec:numerics}

In Papers I and II, we investigated the cosmological evolution and
structure of five DM halos by setting the initial conditions of the
simulations using constrained realizations of Gaussian fields. Our
basic motivation was to perform controlled numerical experiments
designed to study the dependence of the evolution and structure of a
given halo on its merging history. The \citet{HR91} algorithm of
constrained realizations of Gaussian fields has been used to set up
the initial conditions. Papers I and II present the analysis of five
different models of a given DM halo, evolving along various merging
histories. The models were simulated within the framework of the open
CDM (OCDM) model with $\Omega_0=0.3$, $h=0.7$ and $\sigma_8=0.9$,
where $\Omega_0$ is the current cosmological matter density parameter
and $\sigma_8$ is the variance of the density field convolved with a
top-hat window of radius $8\hmpc$ used to normalize the power
spectrum.  This model is very close to the `concordance' $\Lambda$CDM
model in its dynamical properties.  The models are labeled as OCDMa,
OCDMb, etc., following the notations of Papers I and II.  Here we add
a new model, run within the flat-$\Lambda$ cosmology with the
parameters of the WMAP three years data base \citep{wmap3}. The
constraints used to set the model are similar to OCDMa and the model
is labeled as WMAP3a, yet it is performed within a $256^3$
computational box and it starts from a different random realization of
the initial conditions.  A full description of the numerical
simulations, the application of the algorithm of constrained
realizations and the numerical code are given in Papers I and II. The
WMAP3a model is one in a series of runs in the WMAP3 cosmology, to be
reported in a forthcoming paper.  The NFW fitting algorithm is
described in Paper II and the \psd\ power law profile is fitted in a
similar way.

\section{General Considerations}
\label{sec:general}

The density profile of DM halos is well approximated by the NFW profile,
\begin{equation}
\label{eq:NFW}
\rho(r) = {4 \rho_s \over (r/R_s)(1+r/Rs)^2},
\end{equation}
in which the characteristic density ($\rho_s$) and scale radius
($R_s$) define the NFW profile.  Defining the halo as a collection of
particles in a spherical (say) volume in which the mean density equals
some critical over-density (which is in general redshift dependent)
times the mean cosmological density, the virial mass ($M_{vir}$) and
radius ($R_{vir}$) of the halo are determined. It follows that \rvir\
and \rs\ (or equivalently \mvir\ and \ms) are the two independent
parameters that define an NFW halo.

Assuming the NFW parameterization, we can write the \psd\ profile as
\begin{equation}
\label{eq:qnfw}
Q(r) = Q_s  \calQ \bigg( {r \over R_s}\bigg),
\end{equation}
where 
\begin{equation}
\label{eq:calq}
\calQ(x)=  { \rho(x)/\rho_s \over \big[ \sigma(x) / \sigma_s \big]^3},
\end{equation}
\begin{equation}
\label{eq:Qs}
Q_s =  { \rho_s \over  \sigma{_s ^3} },
\end{equation}
$\sigma_s$ is the mean velocity dispersion within $R_s$ and
$x=r/R_s$. To the extent that the DM halos are fitted by the NFW
profile, their $\calQ(x)$ profile should obey a universal
relation. Their cosmological evolution is then determined by the
evolution of $Q_s$.

\section{Evolution of Dark Matter Halos}
\label{sec:results}

\begin{figure*}[!t]
\epsscale{1.0}
\plotone{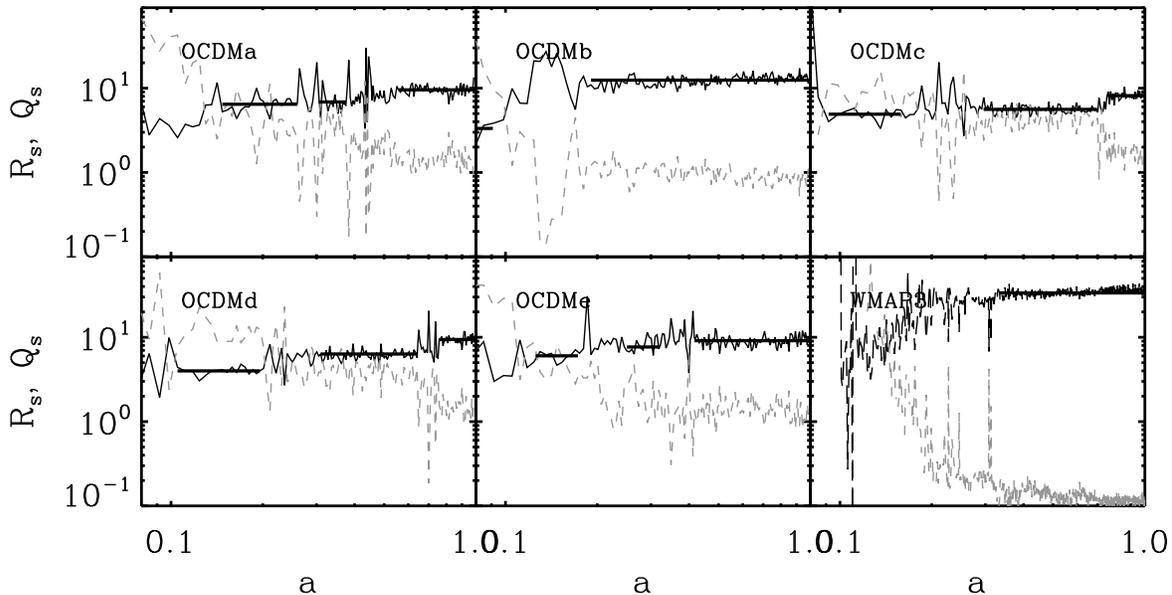}
\caption{The cosmological evolution of the main halo of the six models
is shown as a function of the expansion parameter $a$.  The violent
phases are characterized by sudden increase (decrease) of $R_s$,
indicated by continuous lines, ($Q_s$, dashed lines) and theses
quantities remain roughly constant in the quiescent phase.  At the
violent phases the NFW fitting fails and the resulting $R_s$ and $Q_s$
become unstable. The thick horizontal lines correspond to the mean
value of $R_s$ within the different quiescent phases. The WMAP3 model
exhibits an extended initial phase of a succession of violent mergers
that cannot be resolved into individual ones separated by quiescent
phasess, like in the OCDM models. This is followed by a quiescent
phase. The $Q_s$ curves have been shifted to fit within the window.}
\label{fig:evol}
\end{figure*}

The cosmological evolution of the main halo of the six different
models is best presented by Fig. \ref{fig:evol}, which shows the
evolution of $R_s$ and $Q_s$.  The halo goes through violent episodes
of major mergers and quiescent phases of slow accretion. In the OCDM
models the violent events are well separated by the quiescent phases
characterized by an NFW structure (Papers I and II). The WMAP3 model
goes through an early phase of successive violent mergers, frequent
enough so that the halo does not relax to an NFW-like configuration in
between. This early phase is followed by a quiescent phase, which is
slightly perturbed by mergers not strong enough to be qualified as
major. This is clearly shown by the behavior $R_s$ and $Q_s$, where
$R_s$ ($Q_s$) increases (decreases) discontinuously in the violent
phases and remains constant in the quiescent ones. One should note
that in the violent episodes the halos are not in an equilibrium and
therefore the NFW fitting is very unstable and the resulting $R_s$ and
$Q_s$ parameters are quite erratic. This is reflected by the spiky
behavior of these quantities in the violent phases.

\begin{figure*}[!t]
\epsscale{1.}
\plotone{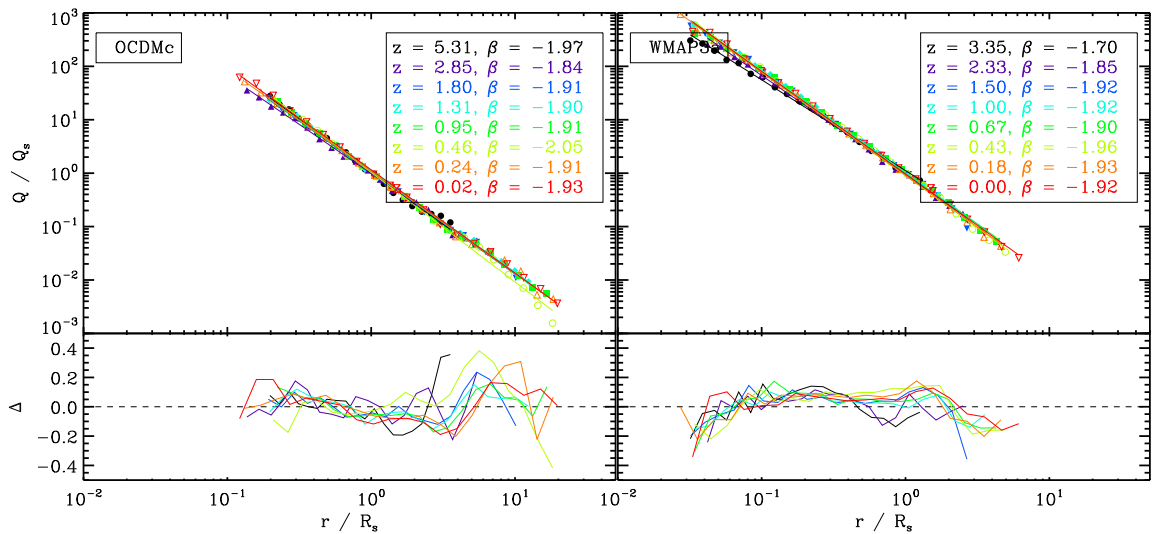}
\caption{The cosmological evolution of the dimensionless phase-space
density profile, $\calQ(r/R_s)$, of the OCDMc (left panel) and the
\LCDM\ WMAP3a (right panel) halos.  The $\calQ(r/R_s)$ profile is
plotted for a sample of snapshots for the two halos.  The value of
$\beta$ for each snapshot is indicated. The bottom panels show the the
fractional residual deviation from a power law. The relative deviation
from a power law behavior is smaller than twenty percent. The
universal form of $\calQ(r/R_s)$ is clearly exhibited. The
$\calQ(r/R_s)$ profiles of the other OCDM models are virtually
identical to the ones shown here.  }
\label{fig:calQ}
\end{figure*}

\begin{figure*}[!t]
\epsscale{1.}
\plotone{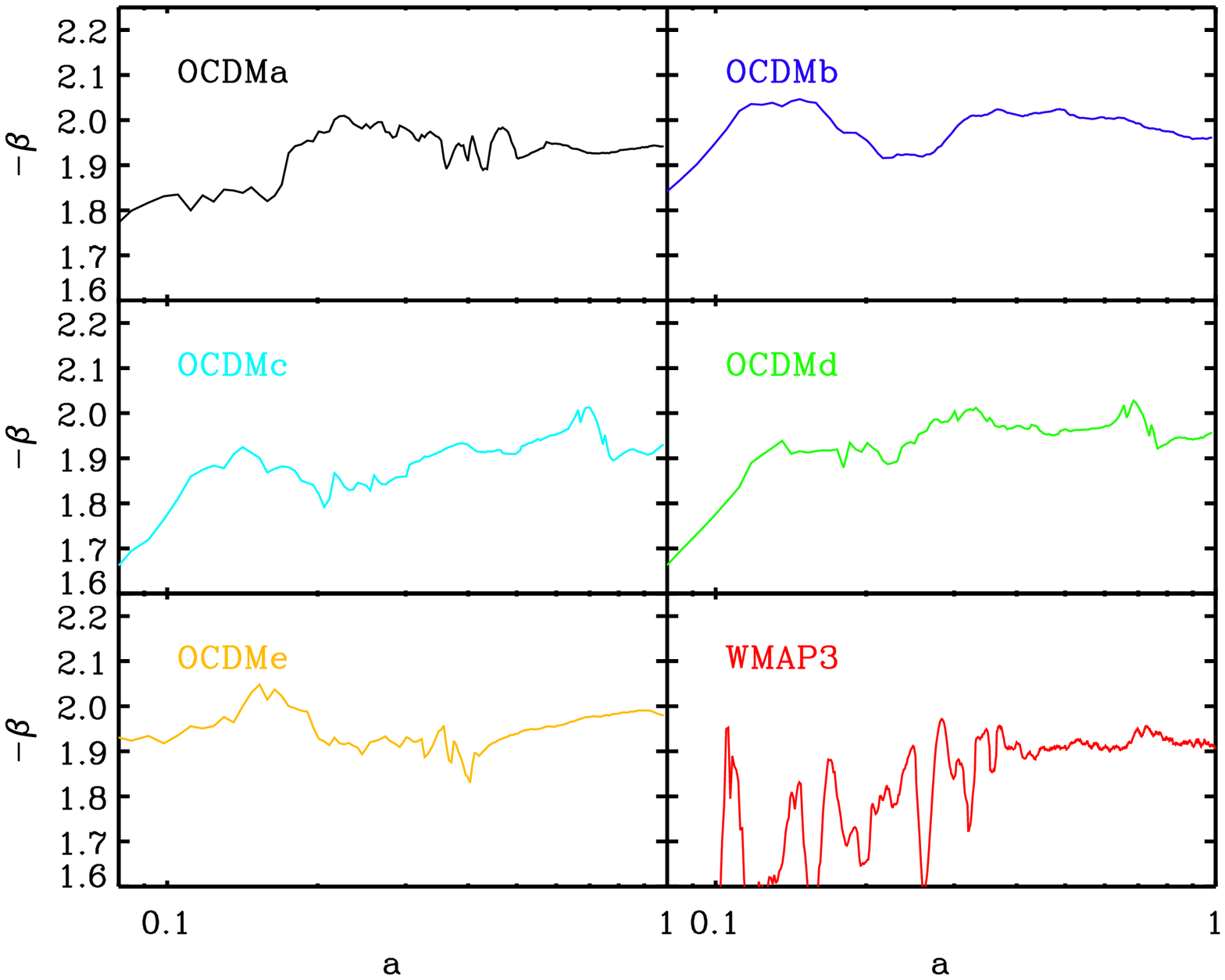}
\caption{ The evolution of the slope of the phase-space density power
law ($\beta$) is plotted against the expansion parameter $a$ for the
six halo models.  }
\label{fig:beta}
\end{figure*}

Evolution of the dimensionless \psd\ profile, $\calQ(x)$, is presented
in Fig. \ref{fig:calQ}, where the $\calQ(r/R_s)$ profiles of the OCDMc
and WMAP3 models are evaluated at different epochs, covering the time
interval from $z=5.3$ ($z=3.35$) for the OCDMc (WMA3) model to the
present epoch.  The evolution of the other models is virtually
identical to the ones shown here. The $\calQ(x)$ profiles are very
closely approximated by a power law, with a fractional deviation of
less than a twenty percent (lower panel of Fig. \ref{fig:calQ}).
Fig. \ref{fig:calQ} shows not only the power law nature of the
profiles but also that indeed the $Q_s$ scaling renders the $\calQ(x)$
profile to a universal time independent power law.  The evolution of
the exponents of the $\calQ(x)$ fitted power law of all the models is
shown in Fig. \ref{fig:beta}.
$\beta$ displays a bumpy behavior, with bumps corresponding to the
major mergers, but generally staying within the $\beta \approx 1.9\pm
0.1$ range. Overall, given the violent character of the halo's
evolution, the robustness of the $\beta$ range is remarkable.

\begin{figure*}[!t]
\epsscale{1.}
\plotone{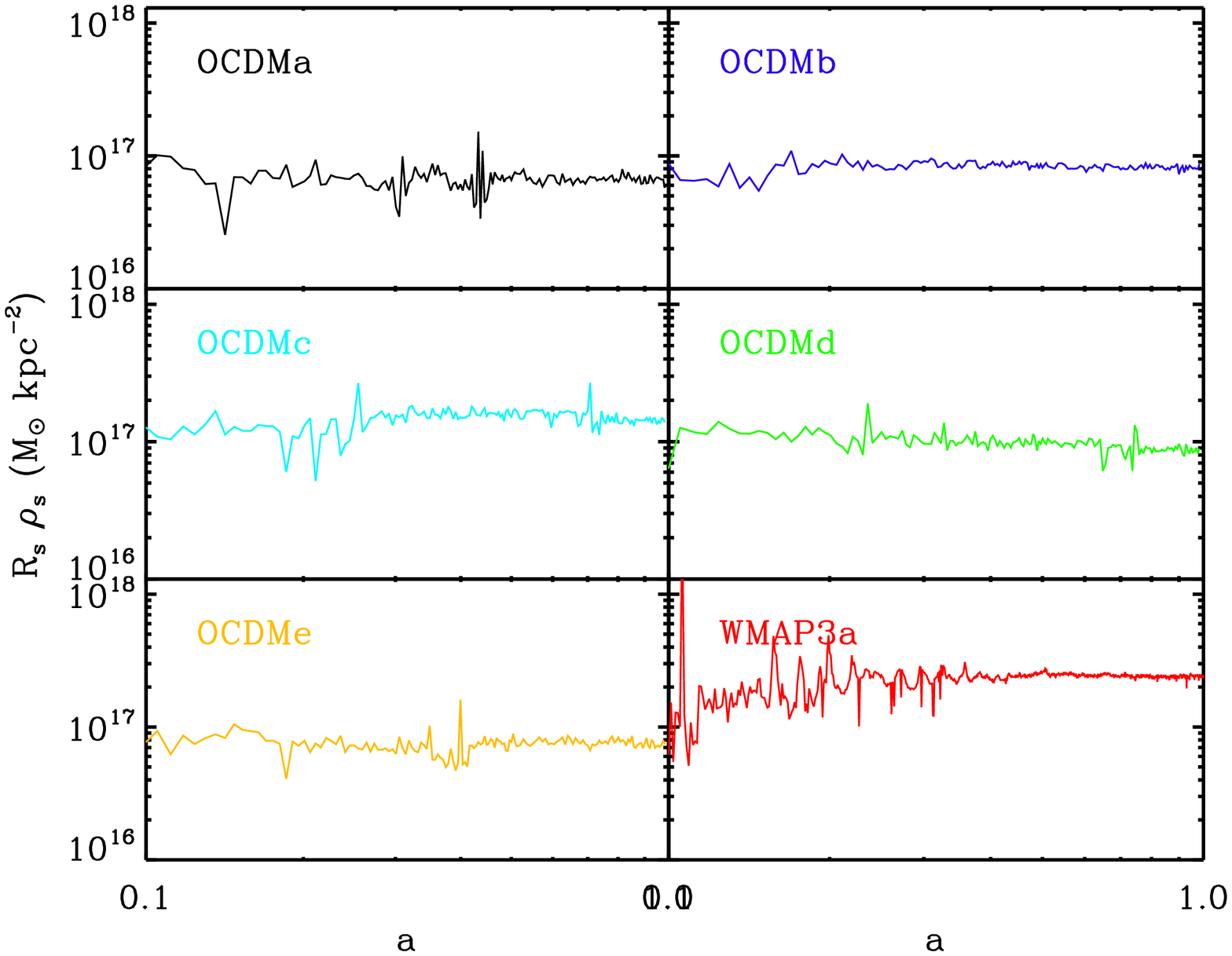}
\caption{ The product $\Sigma_s =\rho_s R_s$ is plotted against the
expansion parameter $a$ for the six halo models.  The panels present
the individual models.  $\Sigma_s$ exhibits a jittery behavior in the
quiescent phases and the large spikes correspond to the violent epochs
in which the NFW fitting is ill-defined. Apart from these, $\Sigma_s$
remains approximately constant throughout the evolution.  }
\label{fig:rhosrs}
\end{figure*}

A close inspection of the halo evolution reveals that the product
$\Sigma_s \equiv \rho_s R_s$ remains approximately invariant as the
halo evolves along the main branch of its merging tree
(Fig. \ref{fig:rhosrs}). The value of $\Sigma_s$ fluctuates around its
mean
value in two different modes..  
undergoes fast and correlated fluctuations of a small amplitude (see
next section and Paper~II). In the violent episodes it exhibits large
deviations from the mean value. This behavior is associated with
timings of major mergers when halos are far from the equilibrium. The
NFW fitting fails here and the resulting NFW parameters are
ill-defined. Ignoring the spikes of the violent phases, and averaging
over the jittery fluctuations in the quiescent phases, $\Sigma_s$
retains its value along the halo evolution. {\it While the invariance
of} $\Sigma_s$ {\it in a given quiescent phase is not surprising, its
ability to retain the value in and across the major merger events is
not obviously expected.}

\section{Self-Similarity and Scaling Relations}
\label{sec:scaling}

Assuming the empirical finding of $\Sigma_s$ invariance, we shall
study its ramifications for the evolution of the phase-space density.
In the quiescent phases, in which the density follows the NFW profile,
a halo is expected to be in virial equilibrium, as corrobrated by
Fig. 13 of Paper~II. In particular the virial ratio, evaluated within
\rs, should have a constant value, namely,
\begin{equation}
\label{eq:vir}
{\sigma{_s^2} \over M_s/R_s} \approx const.
\end{equation}
The virial ratio differs from unity because the inner part of the halo
does not constitute an isolated system, and its value depends on the
shape of the density profile.

\begin{figure*}[!t]
\epsscale{1.}
\plotone{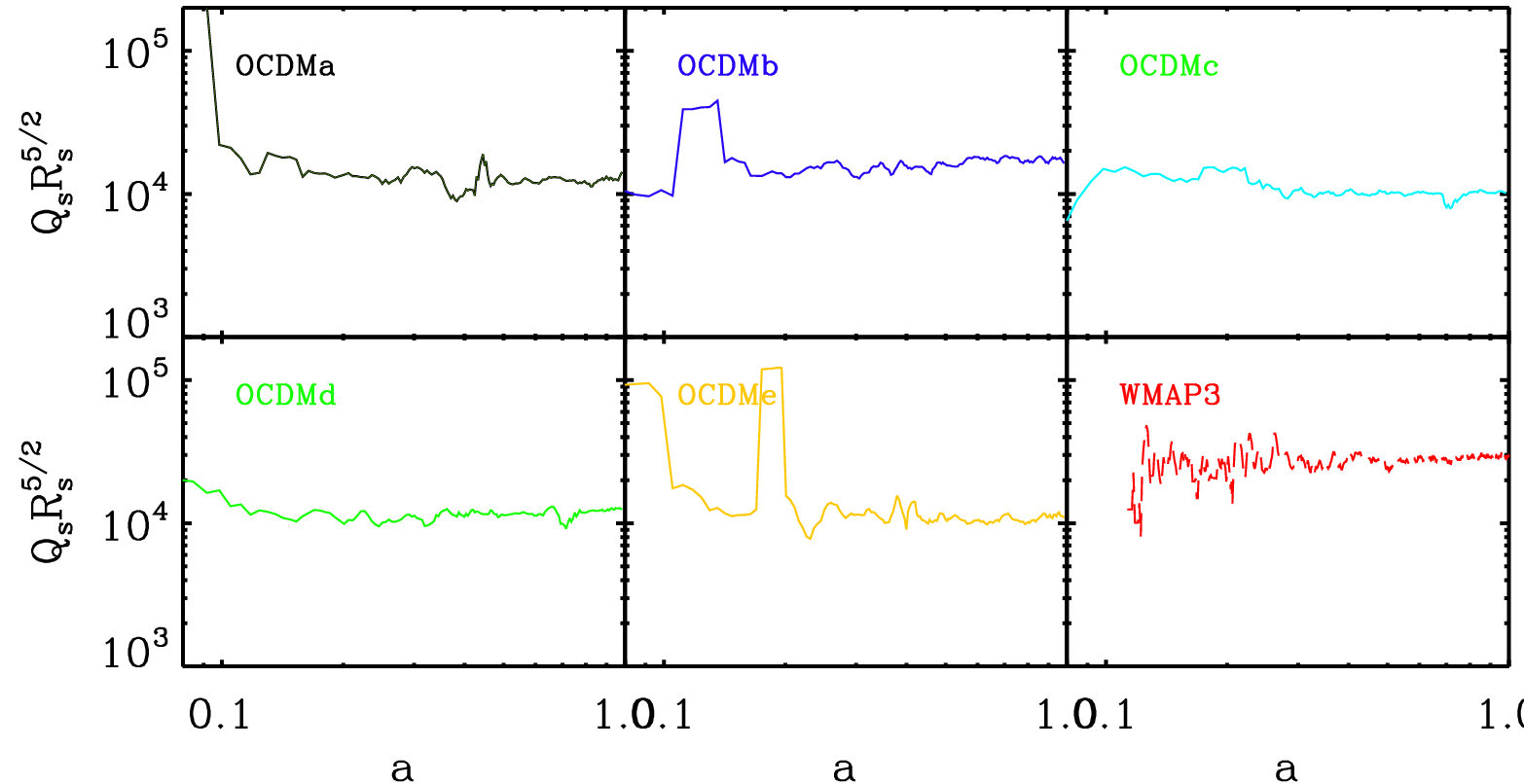}
\caption{ The cosmological evolution $Q_s R{_s^{5/2}}$ of the main
halo of the six models is shown as a function of the expansion
parameter $a$.  $Q_s R{_s^{5/2}}$ behaves very similarly to $\Sigma_s$
(Fig. \ref{fig:rhosrs}).  }
\label{fig:qsra}
\end{figure*}

Assuming the constancy of the virial ratio within $R_s$ and the
invariance of $\Sigma_s$,
\begin{equation}
\label{eq:const}
\sigma{_s^2} \rho_s \propto  {M_s \rho_s \over R_s} \propto \rho{_s^2} R{_s^2} = \Sigma{_s^2}.
\end{equation}
The evolution of the phase-space density profile in a given halo is
described by
\begin{equation}
\label{eq:q-evol}
Q(r) \propto  \Sigma{_s^{-1/2}} R{_s^{-5/2}} \Big({r \over R_s}\Big)^{-\beta}.
\end{equation}
Consequently,   for a given halo  the following invariance   holds:
\begin{equation}
\label{eq:QsRs}
Q_s R{_s^{5/2}} \approx const.
\end{equation}
This prediction has been tested against the five models of Papers I
and II and the \LCDM\ halo.  Fig. \ref{fig:qsra} shows the
cosmological evolution of $Q_s R{_s^{5/2}} $ of these models.  As with
all other quantities that characterize the DM halos, the product $Q_s
R{_s^{5/2}}$ shows a jittery behavior in the quiescent phases and
strong fluctuations in the violent phases, in which the NFW parameters
are ill defined. Apart from this, it remains constant throughout the
evolution.
Only Model C exhibits a small deviation from this invariance. In its
first quiescent phase $Q_s R{_s^{5/2}}$ is larger by a factor of
$\lesssim 1.5$ than its asymptotic value. 

The $\rho_s R_s$ invariance is to be distinguished from the $\rho_s
\propto R{_s^{-m}}$ scaling found in its the low-amplitude jitter
(Paper II) and from $M_s \propto R{_s^\alpha}$ of
\citet{Zhao_etal03}. In Paper II we show that $m \approx 1.39$ in the
last quiescent phases of various halos, while at early times $m
\approx 1.59$. This stands in good agreement with
\citeauthor{Zhao_etal03}'s $\alpha\approx 1.44$ in the ``slow
accretion phase'' and $\alpha\approx 1.92$ in the ``rapid accretion
phase.'' The $m \approx 1.39$ scaling is obtained by analyzing each
quiescent phase, and it reflects the fluctuations of $\rho_s$ and \rs\
around their mean values within that phase. This correlation appear to
be driven by density fluctuations that originate in the region between
the cusp and $R_s$, where the density slope varies between -1 and -2.
The mean values of $\rho_s$ and \rs\ change from one quiescent phase
to another. The $m \approx 1.59$ reflects the variation across the
different phases. The association of the $m \approx 1.59$ and
$\alpha\approx 1.92$ with the early times of the evolution of halos
stems from the fact that the violent phases are more abundant at early
times. The combined analysis of Paper II and the present work yields
the following picture. In the quiescent phases the values of $\rho_s$
and \rs\ fluctuate around constant values, yet their product remains
invariant along the evolution of a halo.

The $\Sigma_s$ invariance implies that $M_s \propto R{_s^2}$.  This is
very close to the $M_s \propto R{_s^{1.92}}$ of \citet{Zhao_etal03}
and the minor discrepancy results from \citeauthor{Zhao_etal03} not
separating explicitly between the quiescent and violent phases. We,
therefore, support and validate one of \citeauthor{Zhao_etal03}'s main
points, namely the $M_s \propto R{_s^\alpha}$ scaling, and set
$\alpha=2$. Thus, we validate also \citeauthor{Zhao_etal03}'s relation
between the evolution of the concentration parameter and the
MAH. Moreover, this relation can be easily extended to the MAH ---
$Q_s$ relation, given the $Q_s R{_s^{5/2}}$ invariance.

A final note concerns the entropy of a system of self-gravitating
collisionless particles.  The definition of the entropy (per particle)
of a monoatomic ideal gas is givein by
\begin{equation}
\label{eq:s}
s=k_B \ln(Q^{-1}) + const,
\end{equation}
where $k_B$ is the Boltzmann constant (\eg\ \citeauthor{dh01}
\citeyear{dh01}; \citeauthor{wn87} \citeyear{wn87}).  Applying this
definition to the DM particles provides one with a formal entropy of
the DM. Our findings concerning structure and evolution of the \psd\
can be easily translated to the language of the entropy of the DM. It
should be noted here that Eq. \ref{eq:s} provides a local measure of
the entropy. The long range nature of the gravitational interactions
prohibits a simple extension of the entropy to an extensive quantity
that characterizes the whole halo. Within this framework we can only
refer to the entropy as a local property.


\section{Discussion and Conclusions}
\label{sec:disc}

The main elements of the structure and evolution of the DM halos can
be summarized as follows.  Halos evolve through two phases, quiescent
and violent ones, which represent the two extreme cases of smooth
accretion and major mergers. In the quiescent phase the halo density
distribution is closely approximated by the NFW profile.  The inner
(within $R_s$) halo mass surface density, $\Sigma_s$, remains
approximately constant throughout its entire evolution.  Most
importantly, the major mergers that take the halo from one quiescent
phase to the other preserve the value of $\Sigma_s$. During the
quiescent phases the halo phase-space density profile follows a power
law of the form $Q(r)=Q_s (r/R_s)^{-\beta}$ with $\beta \approx
1.9$. The cosmological evolution of the phase space density is given
by $Q_s$. Under the invariance of $\Sigma_s$ and the assumption of the
virial equilibrium within $R_s$, the evolution of $Q_s$ is dictated by
$R_s$, so that $Q_s \propto R{_s^{-5/2}}$. In the quiescent phases,
$Q_s$ remains constant and it decreases discontinuously in the violent
phases.
  
The $\Sigma_s$ invariance and the virial theorem (Eq. \ref{eq:const})
imply that the evolution of a halo proceeds while conserving its
surface density and its energy per unit volume, or equivalently the
pressure, within $R_s$. The interesting point is that a typical halo
undergoes a few violent events of major merging that destroy its
equilibrium. Following each event it regains the NFW structure with a
larger $R_s$, while preserving its pressure and the mean surface
density within the new $R_s$.  Summarizing, a DM halo evolves along a
sequence of NFW profiles, with an ever increasing $R_{vir}$ and $R_s$
that grows only discontinuously, in the manner described in Paper I,
while conserving the mean surface density and the pressure within
$R_s$.

The evolution of the $Q(r)$ profile of a given DM halo is predicted by
Eq. \ref{eq:q-evol}. The controlled numerical experiments of Papers I
and II suggest that, for a given halo, $\Sigma_s$ remains constant
over a few quiescent phases intermitted by violent events. All the
halos analyzed here (apart from WMAP3a) emerge from the same
realization of the initial conditions that has been subjected to
different constraints. In many ways these halos can be considered as a
single halo that has been manipulated so as to modify its merging
history. As such they all have roughly the same value of $\Sigma_s$.
The WMAP3a model provides an independent realization of a DM halo in a
different cosmology. The value of $\Sigma_s$ is roughly twice as large
than in all the other model. Of course the extremely poor statistics
of our models cannot teach us about the scatter in $\Sigma_s$.  The
question arises as to what happens in the general case of a large
ensemble of DM halos drawn from a large cosmological simulation. The
scatter in $\Sigma_s$ is expected to determine the evolutionary tracks
of the phase-spase density profiles. It is interesting to study the
possible environment $- \Sigma_s$ correlation, and to what extent this
affects the evolution of the DM halos.

The $\rho_s R_s$ invariance implies that the \citet{Zhao_etal03} $M_s
\propto R{_s^\alpha}$ scaling holds with $\alpha=2$, and so does the
MAH - concentration parameter relation. It follows that this can be
easily extended to calculate a similar MAH $- Q_s$ relation. The
interesting ramification of this result is that any statistical
algorithm for generating DM halos merging trees and/or MAHs can be
extended so as to provide the evolution of the full NFW parameters of
a halo along its merging tree. This can be used in semi-analytical
modeling of galaxy formation in which realizations of merging trees
include the full NFW structure of the halos.

The definition of $Q(r)$ as a phase-space density is only a poor man's
substitute for the 'real' phase-space density in the six-dimensional
phase space. Yet, pushing the analogy further on, a formal entropy is
defined by Eq. \ref{eq:s} as a local variable. This local definition
cannot be extended to provide a global entropy of the halo. It follows
that statements regarding the evolution of the entropy of DM halos are
ill-defined, at least in the current context. Nevertheless, in the NFW
scaling framework, in which the entropy of DM halos is taken as $Q_s$,
it increases with $R_s$. In this context, the quiescent phases with
$R_s\sim $const. correspond to adiabatic processes which preserve the
entropy, and the violent phases --- to non-adiabatic ones in which the
entropy grows.

The method of introducing an entropy by means of Eq. \ref{eq:s} can be
considered as only formal, and its relation to a standard
thermodynamics needs to be questioned. Yet, \citet{fhgy07} have
recently shown that the formally defined entropy of the DM is very
closely associated with the classical (ideal gas) entropy of the
intergalactic gas in clusters of galaxies.  \citeauthor{fhgy07}
studied the entropy of the gas and DM in high resolution adiabatic SPH
cosmological simulations --- `adiabatic' here is used in the sense
that the entropy of the gas can only grow due to shock
waves. Consequently, following the accepted notation, the entropy of
the gas can be expressed by means of $K_{gas}\equiv k_B T/\rho^{2/3}$,
and by analogy $K_{DM} =\sigma{_{DM}^2} / \rho{_{DM}^{2/3}} \propto
Q^{-2/3}$.  The power law behavior of $Q(r)$ implies that
$K_{DM}\propto r^{1.1}$. Interestingly, outside the entropy core of $r
\approx R_s/2$ the gas entropy follows the DM, namely $K_{gas} \propto
K_{DM}$, and the calculated slope agrees well with observations.
Faltenbacher \etal\ amended the definition of $K_{gas}$, such that the
thermal energy is extended to include the kinetic energy of the small
scale (turbulent motion) and the DM and gas densities are normalized
by their mean cosmological values. The newly defined $K_{gas}$ then
coincides with the DM entropy outside the entropy core to a very good
approximation. The analysis of Faltenbacher \etal\ shows that, at
least in galaxy clusters, the DM entropy coincides with the classical,
ideal gas, entropy of the intracluster medium.  It is not clear a
priori why the gas and DM entropies should coincide, but numerical
simulations suggest that they do. The simulations suggest that
observational determination of the entropy of the intracluster gas
should shed light and constrain the behavior of the DM entropy.


\acknowledgments

We thank Ran Rubin for his help with the calculation of the phase
space density profile.  Fruitful discussions with Yuval Birnboim, Adi
Nusser, Noam Soker and Saleem Zaroubi are gratefully
acknowledged. This research has been supported by ISF-143/02 and the
Sheinborn Foundation (to YH), by NASA/LTSA 5-13063, NASA/ATP
NAG5-10823, HST/AR-10284 (to IS), and by NSF/AST 02-06251 (to CH and
IS).




\end{document}